# Good governance and national information transparency: A comparative study of 117 countries

Mahmood Khosrowjerdi[1][0000-0003-1854-1270]

[1] Inland Norway University of Applied Sciences, 2418 Elverum, Norway
mahmood.khosrowjerdi@inn.no

**Abstract.** Information transparency is a major building block of responsible governments. We explored factors influencing the information transparency of 117 world nations. After controlling for the effects of confounding variables of wealth (GDP per capita), corruption rate, population density, human capital, and telecommunication infrastructure, we found that the good governance indices (democracy, economy, and management) were strong and stable predictors of information transparency of world nations.

## 1  Introduction

Transparency (as a general term) is defined as "the quality of being done in an open way without secrets" (Cambridge Dictionary, n.d.), and, the transparency of "a process, situation, or statement" has been termed as "the quality of being easily understood or recognized… in a clear way" (Collins Dictionlary, n.d.).

Although transparency is a prevalent concept in religious texts (See for example, Taufiq, 2015), the academic research on transparency dates to 1980s (Ball, 2009). Transparency have been differently conceptualized and studied in many fields, but three dimensions of *availability*, *quality*, and *clarity* could be common in those conceptions. This is in accordance with the recent perceptions of transparency as "the degree of openness in conveying information" (Ball, 2009). Transparency could be used in different contexts. The transparency of nations/governments could be categorized into two dimensions: 1) the information transparency (the amount, quality, and flow of information in a society) and, 2) the accountability (e.g. fiscal transparency, free media, etc.) (Williams, 2015). The focus of current study is on the first dimension, that is, the information transparency.

Information transparency is a major antecedent of effectiveness of decision-makings of public institutions (Ball, 2009). Information transparency contributes to quality of public services (Bauhr & Carlitz, 2020; Stirton & Lodge, 2001), increases trust in governments (Grimmelikhuijsen, 2012; Song & Lee, 2016), and it is an im-portant perquisite of international relations and collaborations (McCarthy & Fluck, 2017).

The dissemination of credible and fast information is very necessary for citizens of all societies to properly response to global health crises (Annaka, 2021; Arora et al., 2020). A recent study shows that the level of information transparency is strongly associated with the number of COVID-19 death cases of 108 nations (Annaka, 2021).

## 1.1 Factors Influencing Information Transparency of Nations

Previous studies show that many variables could influence the information transparency of nations.

Of the sociodemographic factors, the wealth of the nations (Williams, 2015), the education level of citizens (Lowatcharin & Menifield, 2015), the corruption perception (Rodríguez Bolívar et al., 2013), the technological infrastructures such as access to internet (Alcaraz-Quiles et al., 2014), the governance style (Brusca et al., 2018; Kachouri & Jarboui, 2017; Liu & Liyuan, 2016; Liu et al., 2016), and the population density of nations (Lowatcharin & Menifield, 2015) have been shown as correlators of national information transparency.

Williams (2015) revealed a statistically significant, positive and linear relationship of wealth of countries with the information transparency. The researcher showed that wealthy countries (operationalized by GDP per capita) had higher national information transparency than others. Kumar et al. (2021) investigated the relationships of national cultural values with the e-government development of 78 countries. They revealed the positive effects of wealth (GDP per capita), the positive effects of five cultural dimensions of individualism, uncertainty avoidance, long-term orientation, and indulgence, and the negative effect of power distance, on the e-government development in noted countries.

Transparency can reduce corruption (Lindstedt & Naurin, 2010), and the higher corruption have been prevalent in non-democratic societies with lower transparencies in public institutions or central governments (Rodríguez Bolívar et al., 2013). Previous research shows that the corruption level in a society has a negative relationship with information transparency. Agyei-Mensah (2017) studied the information disclosure behavior of 174 businesses in two countries of Botswana and Ghana and revealed that the firms in countries with lower corruption rate (Botswana) had higher information disclosure rate than businesses in countries with higher corruption rate (Ghana). Similarly, Brusca et al. (2018), as a part of a larger research, explored the relationships of information transparency and corruption level of 75 countries, and found that higher transparency of societies were associated with lower corruption level in noted societies. Furthermore, the researchers revealed that the democratic societies were more likely to have lower corruption and higher transparency level (measured by budget openness of the countries).

Population density have been shown as positive correlators of transparency of nations. In other words, the larger the population of nation, the higher the information needs of citizens of the nation could be. Lowatcharin and Menifield (2015) studied the influences of various governance features of 816 Midwest counties of 12 local governments of USA on their websites' transparency. The researchers found that the total land area (in square miles), the population density (persons per square mile), percentage of minorities (non-white population), education level (percent of population with a bachelor's degree and beyond), and the governance style (council-manager form) were statistically significant predictors of provided online information transparency of counties.

The countries which have enriched their information and communication technologies (ICT), are more likely to provide citizens with the required information, than those who lack those technological infrastructures. For instance, Alcaraz-Quiles et al. (2014) investigated the predictors of transparency of sustainability information of 17 regional governments in Spain, and revealed that of several socioeconomic variables, the population density, access to internet, and education levels of population had positive correlations with the transparency of sustainability information.

The governance style of nation could be a correlator of information transparency. For instance, Guillamón et al. (2011) researched the financial information transparency of 100 largest Spanish municipalities and found that the information transparency was dependent to political factors (ideology), collected taxes by municipalities, and received transfers by municipalities. Explicitly, the researchers found that the left-wing parties were more likely to provide higher transparent information than the right-wing parties.

Rodríguez Bolívar et al. (2013) published a meta-analysis on the predictors of public information disclosures. The researchers summarized, at least, four conclusive results of previous studies. First, the financial situation of public institutions had a positive effect on their information transparency, and this effect was moderated by the type of national governance, that is, the effect of financial situation of firms on information disclosure was higher in institutions of Anglo-Saxon countries which could have efficient governance cultures. Second, the political competitions were a positive predictor of information disclosure. In other words, politicians in national institutions were more likely to be under pressure to disclose information than those working in local institutions. Third, because of getting funds and grants from central governments, the local authorities were more likely to publish public information than central government institutions. Finally, the information transparency was a governance dependent factor, that is, the information transparency was higher in nations with stronger oppositions and political parties. In other words, the higher the national political competitions, the higher the need for information disclosures.

Liu et al. (2016) studied the information disclosure patterns of 516 businesses in Taiwan and found that the type of governance was associated with information transparency of firms. They found that family-owned businesses (which possibly had more autonomy and freedom) were more likely to have higher degrees of information disclosures than other businesses.

Kachouri and Jarboui (2017) investigated the degree of openness of 28 corporates in Tunisia and found that the governance style of corporates (i.e. the "ownership concentration") had negative effect on information transparency of the corporates.

## 1.2 Rationale for this study

The review of current literature shows that the research on information transparency, with a couple of exceptions, has been focused on counties/firms/businesses of a single nation, and the current knowledge is not inclusive for cross-country comparisons.

The previous studies have focused on several types of information transparency (e.g. public, financial, etc.), used numerous measures to assess the transparency (e.g. number of websites, publication of general information, publication of financial information, and so on), used data from various sources, and researched different layers of information transparency (e.g. counties, firms, businesses, regional governments, and so on). Thus, the current literature does not provide us with an inclusive cross-country comparison of information transparency of world nations. This research is going to fill these gaps.

We base our analysis on nations (as a macro layer of information transparency). In our research, the information transparency is operationalized as "the amount, quality, and flow of information in a society and study". We include several variables which have been correlated with information transparency in previous studies, that is, the GDP (per capita), Human Capital, Telecommunication Infrastructure, Corruption Perception, Population Density, and the Governance Style (Democracy, Economy, and Management). In order to have an inclusive research design and to have most of nations in the study, we use the open access data for all nations for year 2008. This would contribute to a general picture of antecedents of information transparency of world nations.

In summary, our research aims to a cross-country analysis of 117 nations which could reveal the antecedents of information transparency of world nations.

## 2 Methods

The study follows quantitative research methodologies, has a secondary data analysis approach, and explores the antecedents/correlators of information transparency of nations. The (open access) data for this study is collected from multiple sources.

After a list-wise alignment of the datasets on included variables in this research, 117 countries were included in the final analyses. The included nations in this study is listed in Appendix 1.

### 2.1 Description of included variables in this study

In this study, we included those variables which were shown as correlators with information transparency of nations.

The description of each included variable in this research, and the data source(s) for noted variable is drawn in Table 1.

The national information transparency data was extracted from Williams (2015) for year 2008. Williams (2015) developed this index based on data from various sources mostly from United Nations, World Bank, and the International Monetary Fund (IMF).

**Table 1. The descriptions of included variables/datasets in this study and their original sources**

| Variable | Description | Source |
|---|---|---|
| GDP per capita | GDP per capita (current, US dollar) | World Bank (2008a) |
| Population Density | A nation-level indicator that shows the population density (The midyear population of a nation divided by land area in square kilometers) of world countries. | World Bank (2008b) |
| Corruption Perception | A nation-level, aggregate index that lists world countries according their score on corruption perception among citizens. The index has a scale of 0-10 (0 equals the highest level of perceived corruption and 10 equals the lowest level of perceived corruption) | Transparency International (2008) |
| Human Capital | A nation-level, aggregate index based on two variables, that is, 1) adult literacy: "the percentage of people aged 15 years and above who can, with understanding, both read and write a short simple statement on their everyday life", and 2) Gross enrolment ratio: "the total number of students enrolled at the primary, secondary and tertiary level, regardless of age, as a percentage of the population of school age for that level". The index has a scale of 0 (the worst score) to1 (the best score). | United nations (UN, 2008) |
| Telecommunication Infrastructure | A nation-level, aggregate index that is composed of five variables: "1) estimated internet users per 100 inhabitants; 2) number of main fixed telephone lines per 100 inhabitants; 3) number of mobile subscribers per 100 inhabitants; 4) number of wireless broadband subscriptions per 100 inhabitants; and 5) number of fixed broadband subscriptions per 100 inhabitants." The index has a scale of 0 (the worst score) to1 (the best score). | United nations (UN, 2008) |
| Governance | *Democracy Status*. A nation-level, aggregate index based on following variables: the state-ness, political participation, rule of law, stability of democratic institutions, and political and social integration. The index has a scale of 0 (the worst score) to1 (the best score). | Bertelsmann Stiftung (2008) |
| | *Economy Status*. A nation-level, aggregate index composed of "the level of socioeconomic development, the organization of market and competition, monetary and fiscal stability, private property, welfare regime, economic performance, and sustainability. The index has a scale of 0 (the worst score) to1 (the best score). | |
| | *Management Status*: This nation-level, aggregate index included subdimensions like the level of difficulty, steering capability, resource efficiency, consensus-building, and international cooperation of nations. The index has a scale of 0 (the worst score) to1 (the best score). | |
| National information transparency | The aggregate data for Information transparency of nations based on three sub-dimensions of information quantity (e.g. amount of social, economic, and financial information available for public), information quality (e.g. the dissemination and disclosure of economic and social information in accordance with international standards), and information infrastructure (e.g. access to information, online information users, etc.). The index has a scale of 0 (the worst score) to100 (the best score). | Williams (2015) * |

Note for Table 1:
* The information transparency data was extracted from noted source for year 2008.

The developed national information transparency index has been used in different academic fields such as politics (e.g. Lührmann et al., 2020), public administration (e.g. Ma & Zheng, 2018; Bauhr et al., 2020), economics (e.g. Goodell & Goyal, 2018; Challe et al., 2019), and media studies (e.g. George, 2018).

## 3 Results

In order to investigate the relationships of included variables in this study with the national information transparency we used Pearson's' correlation analysis.

We explored assumptions of Pearson's' correlation analysis (Schober, Boer, & Schwarte, 2018) based on the level of measurement (i.e. continuous values of included variables in the study), absence of outliers, normality of variables, linearity, and homoscedasticity. The included variables in this study were either ratio or interval, and they were labeled as scale (i.e. continuous). The data had no outliers (−3.29 *standard devistioins* < *the mean value of variables* < +3.29 *standard devistioins*) and the variables had normal distributions (−1 < *skewness* < +1). Linearity and homoscedasticity refer to the shape of the values formed by the scatterplots (see Figure 1). The shape of values formed by the scatterplots were relatively straight line (linearity assumption met) and the distance between the points to that straight line was very close, and relatively tube-like in shape (homoscedasticity assumption met).

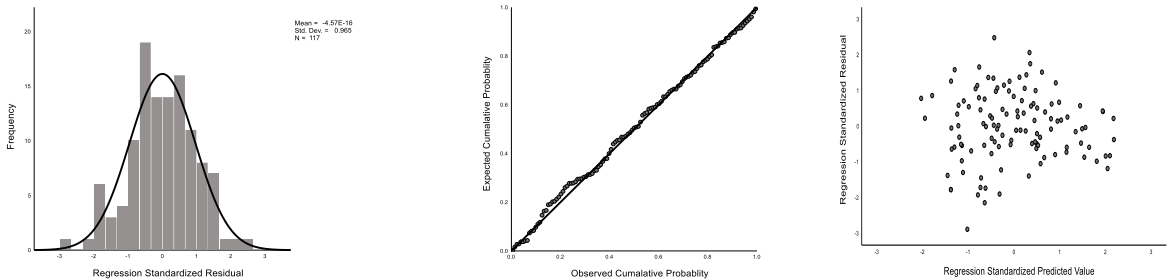

**Fig. 1.** Exploring the assumptions of Pearson's' correlation analysis

Table 2 shows the results of Pearson's correlation analysis. As it is shown, there were positive and significant correlations between GDP (per capita), human capital, telecommunication infrastructure, corruption perception, and the three dimensions of good governance (democracy, economy, and management) with the information transparency of countries. Of the included variables in this study, the three dimensions of good governance have relatively the strongest relationships with the national information transparency.

Because of relatively strong correlations of the three dimensions of good governance with the national information transparency, we calculated the Partial Pearson's correlation coefficients (see Table 3) to explore if those correlations are stable after controlling for the effects of other (confounding variables), that is, GDP (per capita), Human Capital, Telecommunication Infrastructure, Corruption Perception, and the Population Density.

**Table 2.** The Pearson's' correlation coefficients

|  |  | Information Transparency |
|---|---|---:|
| GDP (per capita) | Pearson Correlation | .339** |
|  | Sig. (2-tailed) | .000 |
| Human Capital | Pearson Correlation | .564** |
|  | Sig. (2-tailed) | .000 |
| Telecommunication Infrastructure | Pearson Correlation | .652** |
|  | Sig. (2-tailed) | .000 |
| Corruption Perception | Pearson Correlation | .521** |
|  | Sig. (2-tailed) | .000 |
| Population Density | Pearson Correlation | .101 |
|  | Sig. (2-tailed) | .279 |
| Democracy | Pearson Correlation | .708** |
|  | Sig. (2-tailed) | .000 |
| Economy | Pearson Correlation | .785** |
|  | Sig. (2-tailed) | .000 |
| Management | Pearson Correlation | .618** |
|  | Sig. (2-tailed) | .000 |

Note for Table 2: *Dependent variable*: Information Transparency. *Independent variables*: GDP (per capita), Human Capital, Telecommunication Infrastructure, Corruption Perception, Population Density, Democracy, Economy, and Management.

As it is visible in Table 3, the three dimensions of good governance (democracy, economy, and management) were stable predictor of national information transparency, and the statistically significant correlations of those three dimensions with the national information transparency did not vanished but moderated.

**Table 3.** The Partial Pearson's' correlation coefficients

|  |  | Information Transparency |
|---|---|---:|
| Democracy | Correlation | .527 |
|  | Significance (2-tailed) | .000 |
|  | df | 110 |
| Economy | Correlation | .525 |
|  | Significance (2-tailed) | .000 |
|  | df | 110 |
| Management | Correlation | .437 |
|  | Significance (2-tailed) | .000 |
|  | df | 110 |

Note for Table 3: *Dependent variable*: Information Transparency. *Independent variables*: Democracy, Economy, and Management. *Control variables*: GDP (per capita), Human Capital, Telecommunication Infrastructure, Corruption Perception, and Population Density.

Table 2 and 3 show the statistically significant relationships of some variables with national information transparency, but they do not reveal the nature of those relationships. In order to explore the pattern of relationships of included variables with the national information transparency, the scatterplots of relationships of all variables with the national information transparency are drawn (see Fig 1-8). The scatterplots help to see both the values of individual data points (i.e. nations), and the patterns when the data are taken as a whole (i.e. whether the nations with similar scores on one variable have relatively close scores on national information transparency).

The relationships of wealth of countries (measured by GDP per capita) and the information transparency of nations, which is shown in Fig. 1, do not seem linear. Enough wealth seems a perquisite for good information transparency, but those countries which are among the richest in the world (e.g. Arabic nations) do have a moderate information transparency level.

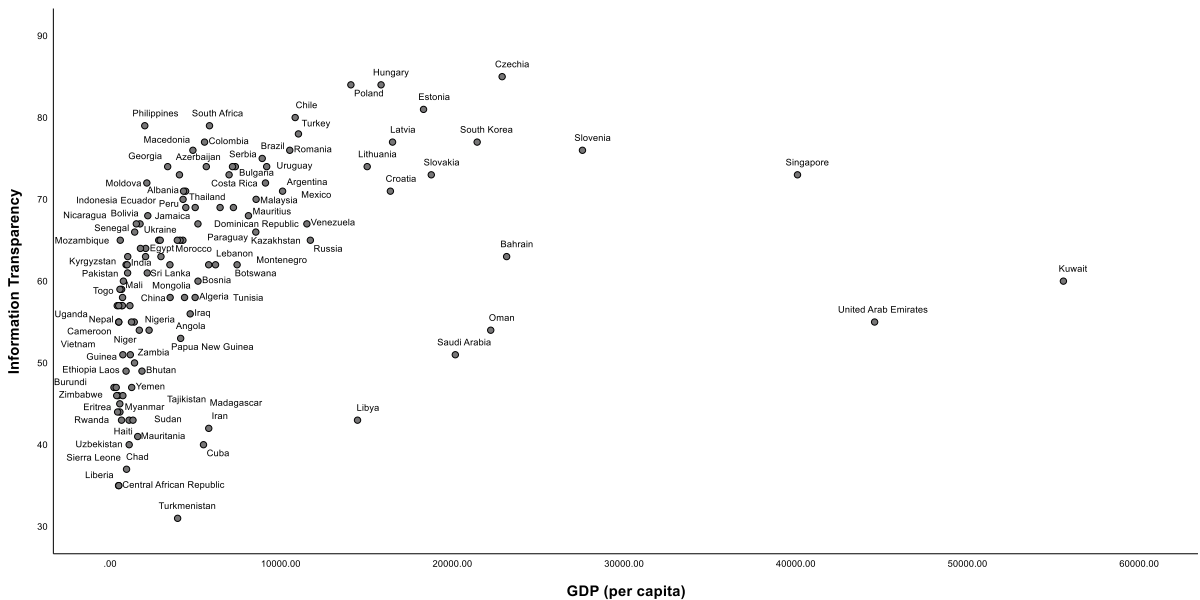

**Fig 1.** Scatterplot of the correlation between GDP (per capita) and the information transparency of nations

Fig. 2 illustrates the relatively strong and positive correlations of human capital (measured by adult literacy and the studentship enrollment ratio) and the information transparency of nations. The figure shows that most of countries which have better human capital are scattered in top-right section of the figure and have higher information transparency scores. However, the previous colonies of Soviet Union (Tajikistan, Uzbekistan, and Turkmenistan) which have relatively high human capital, do have very low information transparency.

The linear associations of telecommunication infrastructure and national information transparency of nations are drawn in Fig. 3. Those nations which have low ICT infrastructure (e.g. Liberia, Chad, Turkmenistan, and so on) are scattered at the bottom-left of the figure and are more likely to have lower information transparency. The nations with excellent ICT infrastructures (e.g. Singapore, South Korea, Estonia,

Czechia, and so on) are grouped at top-right of the figure and have high information transparency too.

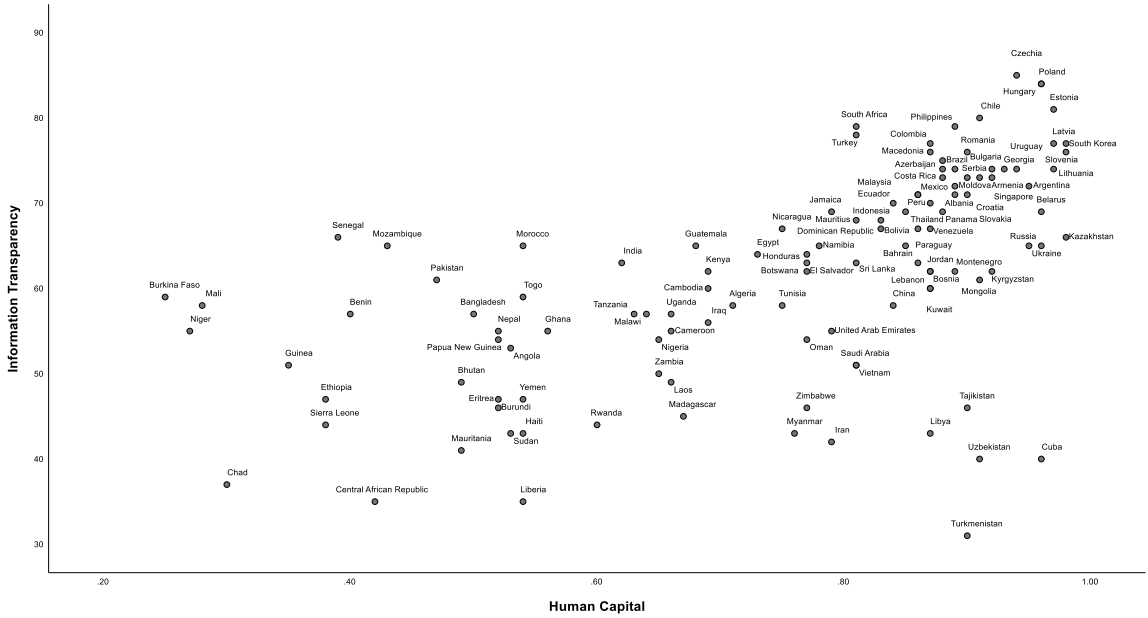

**Fig 2.** Scatterplot of the correlation between human capital and the information transparency of nations

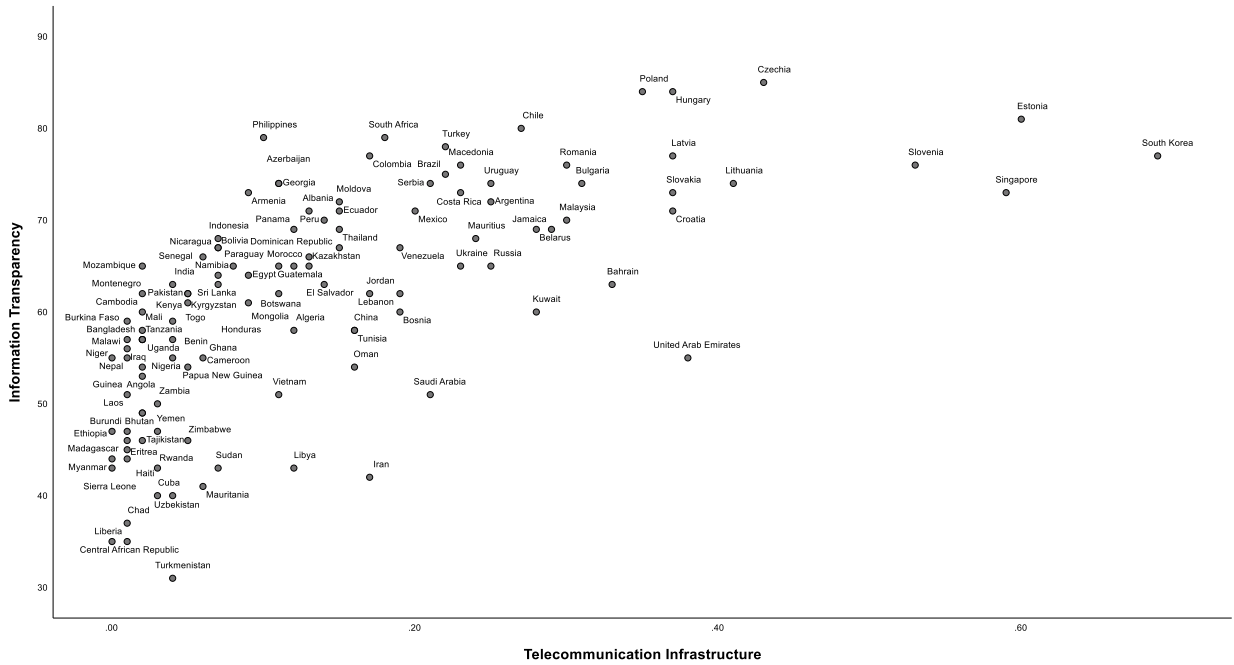

**Fig 3.** Scatterplot of the correlation between telecommunication infrastructure and the information transparency of nations

The relationships of corruption of included nations in this study and their information transparency are scattered in Fig. 4. The figure shows a moderate correlation between mentioned two variables. Countries with high corruption and low information transparency (e.g. Chad, Haiti, Myanmar, Turkmenistan, and Uzbekistan) are scattered in bottom-left of the figure, and the nations with low corruption and high information transparency (e.g. Singapore, Slovenia, and Estonia) are clustered in top-right of the figure.

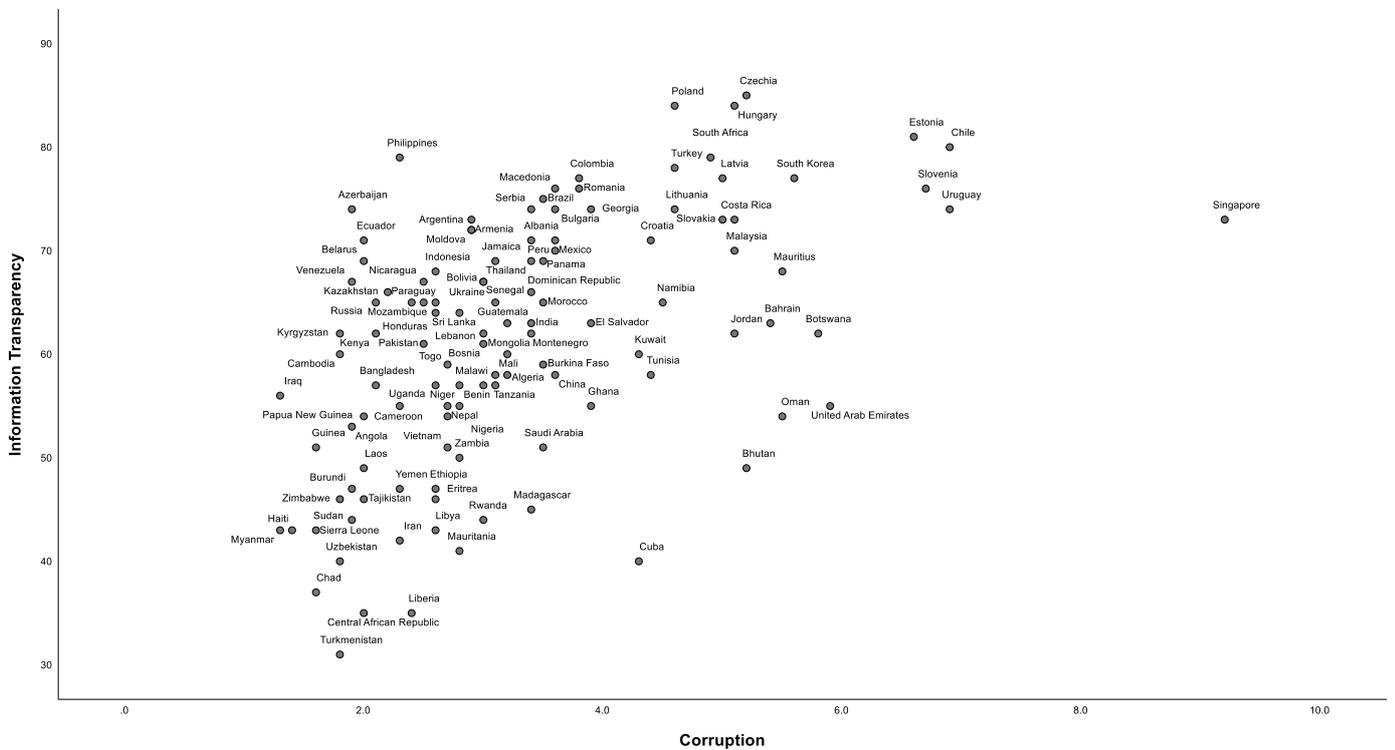

**Fig 4.** Scatterplot of the correlation between corruption perception and the information transparency of nations

Fig. 5 illustrates the positions of including nations in this study according to their scores on population density and information transparency. The relationships of noted variables was not statistically significant, and this is confirmed via the scatterplot too.

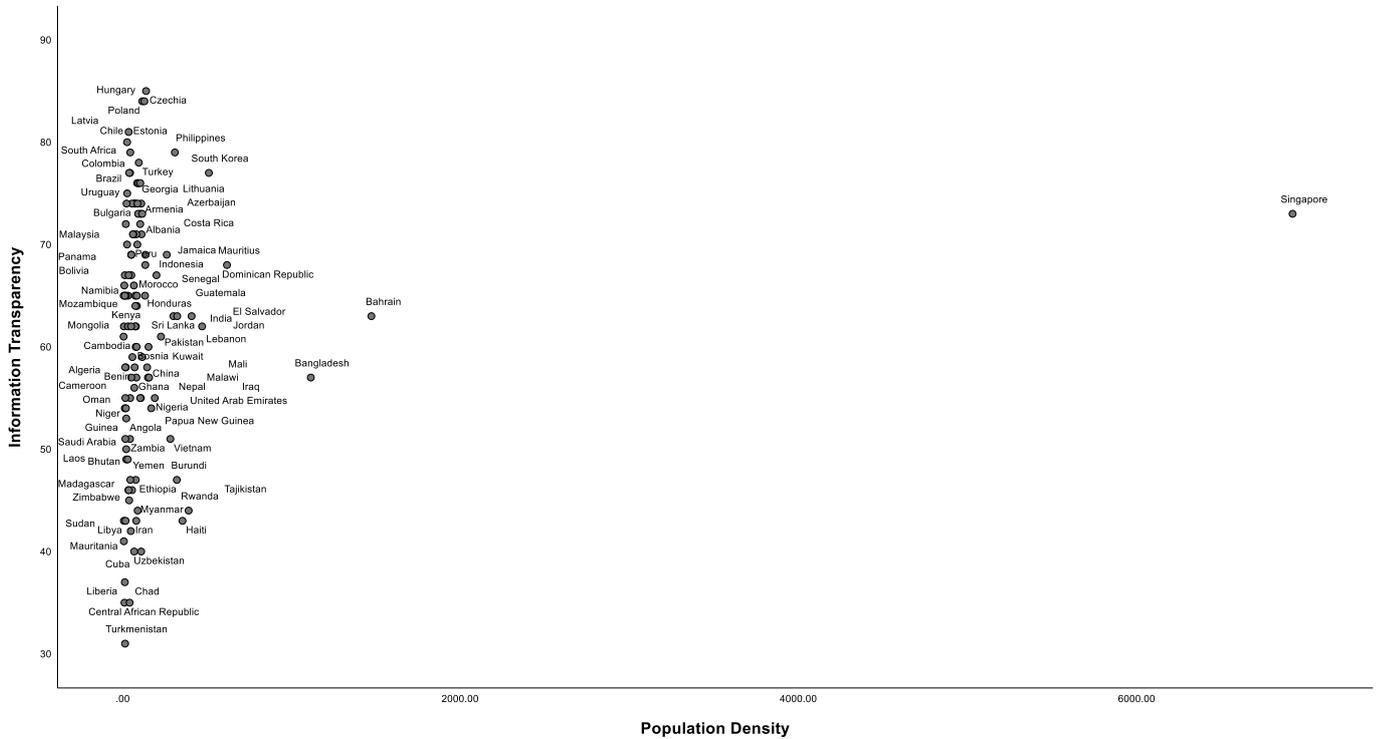

**Fig 5.** Scatterplot of the correlation between population density and the information transparency of nations

Fig. 6 shows the nations' scores on two variables of democracy and national information transparency via a scatterplot. The figure shows a sharp linear relationship of mentioned variables. The democratic nations (those which have highest scores on this dimension) are clustered in top-right of the figure. Nations such as Slovenia, Chechia, Chile, Hungary, and Poland are in this cluster. The non-democratic nations such as Turkmenistan, Chad, and Myanmar are clustered in bottom-left of the figure and they have very high information transparency scores too. This pattern is similar for other countries (e.g. Kyrgyzstan, Russia, Bangladesh, and Sri Lanka) which have average scores on both variables.

Fig. 7 demonstrates the strong linear relationships of two variables of economy and information transparency of nations. Countries such as Singapore, Chile, South Korea, Costa Rica, and all European nations have high scores on both economy and information transparency and are clustered in top-right of the figure, and nations which have low scores on both economy and information transparency (e.g. Liberia, Zimbabwe, and Myanmar) are grouped together in bottom-left of figure.

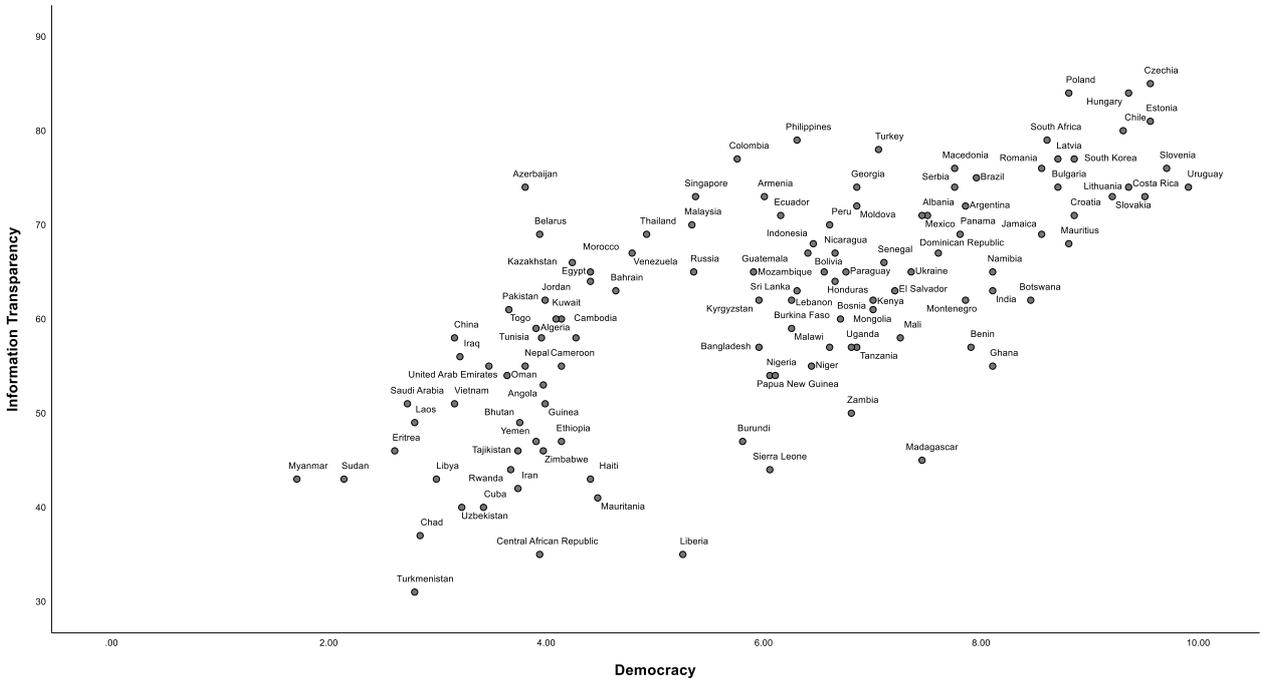

**Fig 6.** Scatterplot of the correlation between democracy and the information transparency of nations

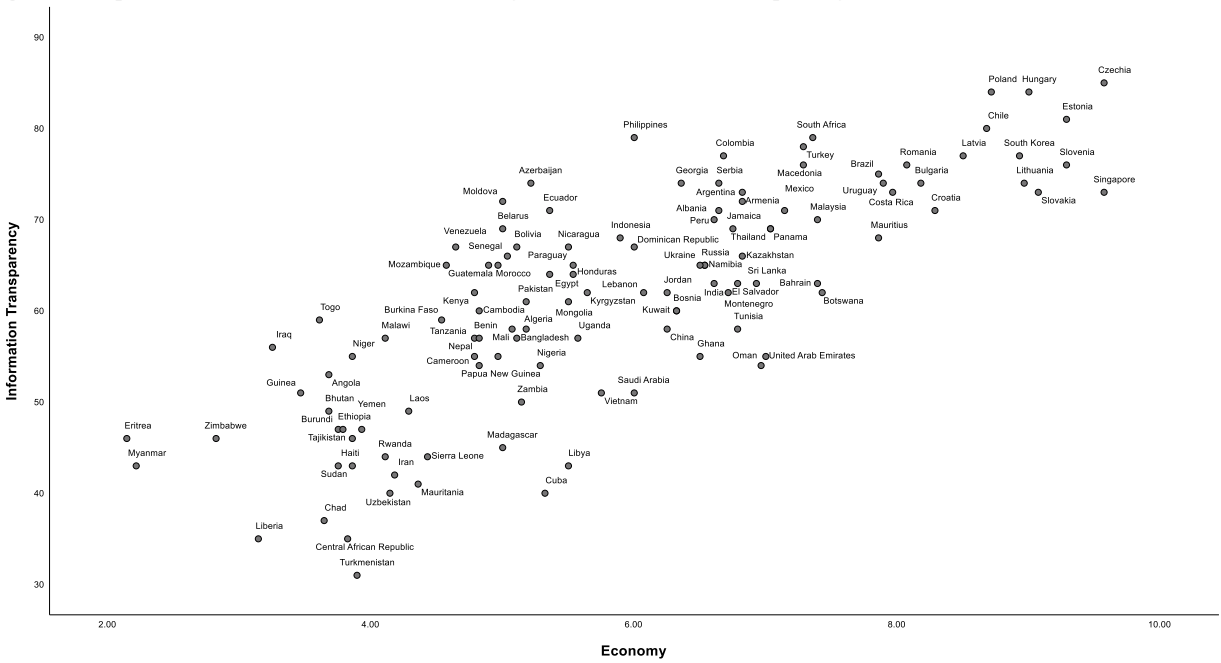

**Fig 7.** Scatterplot of the correlation between economy and the information transparency of nations

The strong links of two variables of management and information transparency of nations are depicted in Fig. 8. The figure shows that those nations which have high scores on both management and information transparency variables (e.g. the European nations, South Africa, South Korea, and Chile) are gathered together in top-right of the figure, and those countries with weak management and low scores on information transparency (e.g. Turkmenistan, Uzbekistan, Chad, and Myanmar) are congregated in bottom-left of the figure.

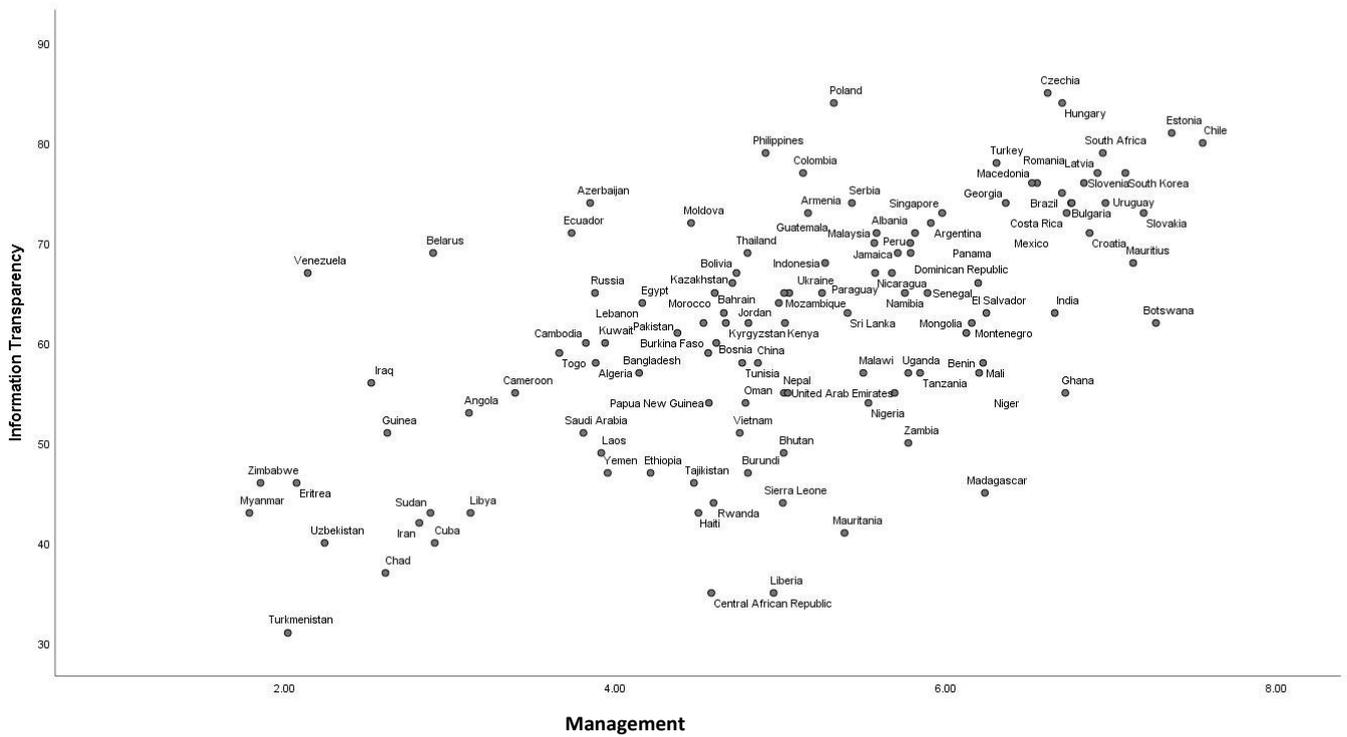

**Fig 8.** Scatterplot of the correlation between management and the information transparency of nations

## 4   Discussions and Conclusion

The aim of this study was to empirically investigate the information transparency of world nations. Our findings provided evidence that seven variables of GDP (per capita), human capital, telecommunication infrastructure, corruption perception, democracy, economy, and management are important correlators of the information transparency of nations.

The wealth of countries (as operationalized by GDP per capita in this study) had a positive and linear relationships with the national information transparency. It means that enough money could be regarded as a perquisite to deliver transparent information to citizens, and the transparency of information could flourish the national economy.

This is in accordance with previous findings of Williams (2015) which reveals that the wealthy nations were more likely to have better information transparency, or the results of studies which highlight the information transparency as a positive predictor of economic performance (e.g. Cimpoeru & Cimpoeru, 2015).

However, wealth could be an antecedent for many other variables in this study, for instance, telecommunication infrastructures and human capital.

The relationships of human capital (measured by literacy and studentship ratio) with the national information transparency was statistically significant in this study. In other words, the literacy of citizens plays an important role in disseminating transparent information. The transparency of governments does not automatically generate informed citizens. The citizens should have enough literacy to communicate with the authorities, and to understand/give feedback on the delivered information to them. As correctly highlighted by Lindstedt and Naurin (2010), "the reforms focusing on increasing transparency should be accompanied by measures for strengthening citizens' capacity to act upon the available information".

The strong correlations of telecommunication infrastructure with information transparency of nations in this study is in accordance with the previous findings of researchers. For example, Alcaraz-Quiles et al. (2014) showed that those local governments which had better technological infrastructures were more likely to be more transparent in delivering information to citizens. The Public's access to information is a hinderance for corruption too (DiRienzo et al., 2007).

Our findings show that the corruption level in societies have negative effects on information transparency. In corrupted societies, the information transparency could be regarded as disclosure risk for authorities, and the information and communication channels would be more controlled. In other words, the ICT could provide citizens a tool for monitoring the process, and finally, could result in reduced corruption of public institutions (DiRienzo et al., 2007).

The last major finding of this study is the strong relationships of good governance (democracy, economy, and management) with the information transparency of nations. The statistical analyses in this study revealed that the good governance dimen-sions was a stable predictor of national information transparency and this association was statistically significant after controlling for the effects of five confounding variables, that is, the GDP (per capita), human capital, telecommunication infrastructure, corruption perception, and population density. The three indicators of good governance, that is, economy status, governance status, and democracy status were among the strongest predictors of national information transparency. This shows to somewhat that information transparency is a cultural-dependent variable. In democratic societies, the trust is accelerator of using public information systems (Pérez-Morote et al., 2020), the governments are more responsible, and the citizens are very informed and engaging (Milner, 2002).

This research has limitations that should be noted. First, the included nations in this research is wide-ranging but not complete. Some countries (e.g. USA and UK) are not included in this study because the data for at least one of investigated variables in this study was not available for them. Second, from the methodological point of view, the included variables in this study are not inclusive. In could exist many other relevant variables which could influence information transparency of nations. For example, it would be interesting to see whether feminine societies which are characterized by high concerns with the quality of life would have better positions in information transparency indices. Another important variable which is not included in this study is the national culture. There is evidence (e.g. Gong et al., 2007; Khosrowjerdi et al., 2020) that cultural values could influence information access and use of citizens, and it could possibly influence the information transparency of world nations too.

**Appendix 1.** Included nations in this study

Albania, Algeria, Angola, Argentina, Armenia, Azerbaijan, Bahrain, Bangladesh, Belarus, Benin, Bhutan, Bolivia, Bosnia, Botswana, Brazil, Bulgaria, Burkina Faso, Burundi, Cambodia, Cameroon, Central African Republic, Chad, Chile, China, Colombia, Costa Rica, Croatia, Cuba, Czechia, Dominican Republic, Ecuador, Egypt, El Salvador, Eritrea, Estonia, Ethiopia, Georgia, Ghana, Guatemala, Guinea, Haiti, Honduras, Hungary, India, Indonesia, Iran, Iraq, Jamaica, Jordan, Kazakhstan, Kenya, Kuwait, Kyrgyzstan, Laos, Latvia, Lebanon, Liberia, Libya, Lithuania, Macedonia, Madagascar, Malawi, Malaysia, Mali, Mauritania, Mauritius, Mexico, Moldova, Mongolia, Montenegro, Morocco, Mozambique, Myanmar, Namibia, Nepal, Nicaragua, Niger, Nigeria, Oman, Pakistan, Panama, Papua New Guinea, Paraguay, Peru, Philippines, Poland, Romania, Russia, Rwanda, Saudi Arabia, Senegal, Serbia, Sierra Leone, Singapore, Slovakia, Slovenia, South Africa, South Korea, Sri Lanka, Sudan, Tajikistan, Tanzania, Thailand, Togo, Tunisia, Turkey, Turkmenistan, Uganda, Ukraine, United Arab Emirates, Uruguay, Uzbekistan, Venezuela, Vietnam, Yemen, Zambia, Zimbabwe